\documentclass[a4paper]{amsart}
\date{October 22, 2020}
\def\arsinh{\mathfrak{Arsin}}
\def\bS{{\mathbb{S}^2}}
\def\rz{\mathbb{R}}
\def\cE{\mathcal{E}}
\def\ge{\mathfrak{e}}

\def\rd{\mathrm{d}}
\def\cE{\mathcal{E}}
\def\cH{\mathcal{H}}
\def\cTg{{{\mathcal{T}}_>}}
\def\D{\mathcal{D}}
\def\tf{T^\mathrm{TF}}
\def\TF{\mathcal{T}^\mathrm{TF}}
\def\TFW{\mathcal{E}_Z^\mathrm{TFWD}}
\def\V{\mathcal{V}}
\def\W{\mathcal{T}^\mathrm{W}}
\def\x{X}
\def\X{\mathcal{X}}
\def\gtf{{\gamma_\mathrm{TF}}}

\def\supp{\mathrm{supp}}
\newtheorem{satz}{Theorem}

\title[Statistical Theory of Heavy Atoms]{A Statistical Theory of
  Heavy Atoms:\\ Energy and Excess Charge}
\begin{document}
\author[H. Chen]{Hongshuo Chen} \address{College of
  Mathematics and Statistics, Chongqing University, Chongqing 401331,
  China} \email{hongshuo.chen@gmail.com}

\author[R. Frank]{Rupert L. Frank} \address{Mathematisches Institut,
  Ludwig-Maximilans Universit\"at M\"unchen, Theresienstr. 39, 80333
  M\"unchen, Germany, and Munich Center for Quantum Science and
  Technology (MCQST), Schellingstr. 4, 80799 M\"unchen, Germany, and
  Mathematics 253-37, Caltech, Pasadena,
  CA 91125, USA} \email{rlfrank@caltech.edu}

\author[H. Siedentop]{Heinz Siedentop} \address{Mathematisches
  Institut\\ Ludwig-Maximilans Universit\"at M\"unchen,
  Theresienstra\ss e 39\\ 80333 M\"unchen\\ Germany\\ and Munich Center for
  Quantum Science and Technology (MCQST)\\ Schellingstr. 4\\ 80799
  M\"unchen, Germany} \email{h.s@lmu.de}

\maketitle

\begin{abstract}
  The purpose of this note is to give an elementary derivation of a
  lower bound on the relativistic Thomas-Fermi-Weizs\"acker-Dirac
  functional of Thomas-Fermi type and to apply it to get an upper
  bound on the excess charge of this model.
\end{abstract}

\section{Introduction \label{Einleitung}}

The description of heavy atoms suffered for a long time from the fact
that the naive adaptation of Thomas-Fermi to the relativistic setting
leads to a functional that is unbounded from below (see Gombas
\cite[\S 14]{Gombas1949} and \cite[Chapter III, Section
16.]{Gombas1956} for reviews). As late as 1987 Engel and Dreizler
\cite{EngelDreizler1987} solved this problem deriving a relativistic
Thomas-Fermi-Weizs\"acker-Dirac functional $\TFW$ from quantum
electrodynamics. For atoms of atomic number $Z$ and electron density
$\rho$ and velocity of light $c$ the functional, written in Hartree
units, is
\begin{equation}
\label{TFWD}
\TFW(\rho)
:= \W(\rho)+\TF(\rho)-\X(\rho) +\V(\rho). 
\end{equation}
The first summand on the right is an inhomogeneity correction of the
kinetic energy generalizing the Weizs\"acker correction.  Using the
abbreviation $p(x):= (3\pi^2 \rho(x))^{1/3}$,
\begin{equation}
  \label{W}
  \W(\rho):=
  \int_{\rz^3}\rd x{3\lambda\over8\pi^2}(\nabla p(x))^2c\,f(p(x)/c)^2
\end{equation}
with $ f(t)^2:=t(t^2+1)^{-\frac12}+2t^2(t^2+1)^{-1}\arsinh(t)$ where
$\arsinh$ is the inverse function of the hyperbolic sine and
$\lambda\in\rz_+$ is given by the gradient expansion as $1/9$ but in
the non-relativistic analogue sometimes taken as an adjustable
parameter (Weizs\"acker \cite{Weizsacker1935}, Yonei and Tomishima
\cite{YoneiTomishima1965}, Lieb \cite{LiebLiberman1982,Lieb1982A}).
The second summand is the relativistic generalization of the
Thomas-Fermi kinetic energy. It is
\begin{equation}
  \label{TF}
  \TF(\rho):=\int_{\rz^3}\rd x{c^5\over8\pi^2} \tf(\tfrac{p(x)}c) 
\end{equation}
with
$\tf(t):=t(t^2+1)^{3/2}+t^3(t^2+1)^{1/2}-\arsinh(t)-{8\over3}t^3$.
The third summand is a relativistic generalization of the exchange
energy. It is
\begin{equation}
  \label{X}
  \X(\rho):= \int_{\rz^3}\rd x{c^4\over8\pi^3} \x(\tfrac{p(x)}c)
\end{equation}
with $\x(t):= 2t^4-3[t(t^2+1)^\frac12-\arsinh(t)]^2$, and, eventually,
the last summand is the potential energy, namely the sum of the
electron-nucleus energy and the electron-electron energy. It is
\begin{equation}
  \V(\rho):= -Z\int_{\rz^3}\rd x\rho(x)|x|^{-1}+\underbrace{\tfrac12\int_{\rz^3}\rd x \int_{\rz^3}\rd y\rho(x)\rho(y)|x-y|^{-1}}_{=:\D[\rho]}.
\end{equation}

We note that, as $c\to\infty$, all integrands of $\TFW$ tend pointwise
to the corresponding part of the non-relativistic
Thomas-Fermi-Weizs\"acker-Dirac functional
$$\cE_Z^\mathrm{nr}(\rho) = \int_{\rz^3}\rd x\left( \tfrac\lambda2 |\nabla\sqrt\rho(x)|^2+\tfrac3{10}\gtf\rho(x)^\frac53 -\tfrac34(\tfrac3\pi)^\frac13\rho(x)^\frac43-\tfrac Z{|x|}\rho(x)\right) +\D[\rho]$$
with $\gtf:=(3\pi^2)^\frac23$ suggesting that we might expect a lower
bound of Thomas-Fermi type when $c$ is large. We will prove in Section
\ref{ua} that this is indeed true. The bound will allow us to
implement the method of \cite{Franketal2018} in the present context
and give an improved bound on atomic excess charges. This is carried
through in Section \ref{Ueberschuss}.

\section{Bound on the Energy \label{ua}}

\subsection{The Domain of $\TFW$}
First we discuss the domain of the functional. To this end, we write
$F(t):= \int_0^tf(s)\rd s$ for the antiderivative of $f$. Then
\begin{equation}
  \label{WF}
  \W(\rho) ={3\lambda c^3\over8\pi^2} \int_{\rz^3}\rd x |\nabla (F\circ (p/c) (x))|^2.
\end{equation}
This allows to define $\TFW$ on
\begin{equation}
  \label{domain}
  P:=\{\rho\in L^\frac43(\rz^3)|\rho\geq0,\ \D[\rho]<\infty, F\circ p\in D^1(\rz^3)\}.
\end{equation}

\subsection{Lower Bound}

We turn to the lower bound itself and address the parts separately.

\subsubsection{The Weizs\"acker Energy}
Since $F(t)\geq t\sqrt{\arsinh(t)}/2$ (see \cite[Formula
(90)]{ChenSiedentop2020}), Hardy's inequality gives the lower bound
\begin{equation}
  \label{usW}
  \W(\rho)
  \geq {3\lambda c\over2^7\pi^2} \int_{\rz^3}\rd x
  {p(x)^2\arsinh(p(x)/c)\over|x|^2}={3^\frac53\lambda c\over2^7\pi^\frac23} \underbrace{\int_{\rz^3}\rd x
  {\rho(x)^\frac23\arsinh(\tfrac{p(x)}c)\over|x|^2}}_{=:\cH(\rho)}.
\end{equation}

\subsubsection{The Potential Energy}
Pick a density $\sigma\in P$ of finite mass and set
$\varphi_\sigma:=Z|\cdot|^{-1}-\sigma*|\cdot|^{-1}$.  Since $\sigma$
is nonnegative, we have $\varphi_\sigma(x)\leq Z/|x|$. Then
\begin{equation}
  \label{ww}
  \V(\rho)= -\int_{\rz^3}\rd x \varphi_\sigma(x)\rho(x) -2\D(\sigma,\rho)+\D[\rho] 
  \geq -\int_{\rz^3}\rd x \varphi_\sigma(x)\rho(x) - \D[\sigma].
\end{equation}
Splitting the integrals at $s$, using \eqref{ww}, and Schwarz's
inequality yields
\begin{equation}
  \label{P}
  \begin{split}
    \V(\rho)
    \geq& -\int_{p(x)/c<s}\rd x\varphi_\sigma(x)\rho(x)\\
    &-Z\int_{p(x)/c\geq s}\rd x {\rho(x)^\frac13\over|x|}
    \arsinh(\tfrac{p(x)}c)^\frac12
    {\rho(x)^\frac23\over \arsinh(\tfrac{p(x)}c)^\frac12}-\D[\sigma]\\
    \geq&
    -{Z\over\arsinh(s)^\frac12}\cH(\rho)^\frac12\cTg(\rho)^\frac12
    -\int_{p(x)/c<s}\rd x\varphi_\sigma(x)\rho(x) -\D[\sigma]
    \end{split}
\end{equation}
with $\cTg(\rho):= \int_{p(x)/c>s}\rd x \rho(x)^\frac43 $.

\subsubsection{The Thomas-Fermi Term}
First, we note that 
\begin{equation}
  \label{tfung1}
  \rz_+\to\rz_+,\ t\mapsto\tf(t)/t^5,
\end{equation}
is strictly monotone decreasing from $4/5$ to $0$ and
\begin{equation}
  \label{tfung2}
  \rz_+\to\rz_+,\ t\mapsto\tf(t)/t^4,
\end{equation}
is strictly increasing from $0$ to $2$. Thus
\begin{equation}
  \label{tfl}
  \begin{split}
    &\TF(\rho)=\int_{p(x)/c<s}\rd x\frac{c^5}{8\pi^2}\tf(\tfrac{p(x)}c)+ \int_{p(x)/c\geq s}\rd x \frac{c^5}{8\pi^2}\tf(\tfrac{p(x)}c)\\
    \geq&\int_{p(x)/c<s}\rd x{3\over10}{5\tf(s)\over 4s^5}\gtf\rho(x)^\frac53+ \int_{p(x)/c\geq s}\rd x \frac{\tf(s)}{s^4}\tfrac38(3\pi^2)^\frac13c\rho(x)^\frac43\\
    =&\frac3{10}{5\tf(s)\over 4s^5}\gtf\int_{p(x)/c<s}\rd x \rho(x)^\frac53+\frac38\frac{\tf(s)}{s^4}\gtf^\frac12c\cTg(\rho).
  \end{split}
\end{equation}

\subsubsection{Exchange Energy}
Since $X$ is bounded from above and $X(t)=O(t^4)$ at $t=0$, we have
that for every $\alpha\in[0,4]$ there is a $\xi_0$ such that
$X(t)\leq \xi_0 t^\alpha$. We pick $\alpha=3$ in which case
$\xi_0\approx 1.15$. Thus
\begin{equation}
  \label{xl}
  \X(\rho)\leq {c\xi_0\over4\pi}N = \xi c N.
\end{equation}
with $\xi:=\xi_0/(4\pi)\approx 0.0914$.

\subsubsection{The Total Energy\label{ax}}
Adding everything up yields
\begin{equation}
  \label{hardyww90}
  \begin{split}
    \TFW(\rho)
    \geq &{3^\frac53\lambda c\over2^7\pi^\frac23} \cH(\rho)
    +\frac38\frac{\tf(s)}{s^4}\gtf^\frac12c\cTg(\rho)
    -{Z\over \arsinh(s)^\frac12}\cH(\rho)^\frac12\cTg(\rho)^\frac12\\
    & +
    \int_{\frac{p(x)}c<s}\rd x(\tfrac3{10}\underbrace{5\tfrac{\tf(s)}{4s^5}\gtf}_{=:\gamma_e(s)}\rho(x)^\frac53-\varphi_\sigma(x)\rho(x))
    -\D[\sigma] -\xi c N.
  \end{split}
\end{equation}
We pick $s\in\rz_+$ such that the sum of the first three summands of
\eqref{hardyww90} is a complete square, i.e., fulfilling
\begin{equation}
  \label{vollquad}
  \sqrt{{3^\frac53\over2^7\pi^\frac23}{3\tf(s)(3\pi^2)^\frac13\over8s^4}}
  ={Z\over c\sqrt\lambda}{1\over2\arsinh(s)^\frac12}.
\end{equation}
The solution is uniquely determined, since $\tf(s)/s^4$ is strictly
monotone increasing from 0 to 2 and $\arsinh(s)$ is also monotone
increasing from $0$ to $\infty$. Call the corresponding $s$
$s_0$. Obviously, $s_0$ depends only on $\kappa:=Z/(c\sqrt{\lambda})$
and is strictly monotone increasing from $0$ to $\infty$.

Eventually we pick $\sigma(x):=\rho(x)\theta(sc-p(x))$. Summing the
first three terms of the second line of \eqref{hardyww90} yields the
Thomas-Fermi functional with Thomas-Fermi constant $\gamma_e(s_0)$
evaluated at $\sigma$. Minimizing this functional and scaling in
$\gamma$ yields
\begin{equation}
  \label{resultat}
  \TFW(\rho)\geq  -\tfrac{4s_0^5}{5T^\mathrm{TF}(s_0)} e^\mathrm{TF}Z^\frac73 - \xi c N
\end{equation}
where $-e^\mathrm{TF}$ is the Thomas-Fermi energy of hydrogen (with
the physical value of the Thomas-Fermi constant, namely $\gtf$).

The function $s_0$ tends exponentially to $\infty$ as
$\kappa\to\infty$. Thus \eqref{resultat} is merely an exponential
lower bound for large $Z$ and fixed $\lambda$ and $c$. However, if we
fix $\kappa\in\rz_+$, then we have a Thomas-Fermi type lower bound
with a correction term linear in $c N$. In conclusion we have
\begin{satz}
  \label{satz1}
  For given $c,\lambda,Z\in\rz_+$, set
  $\kappa:=Z/(c\sqrt\lambda)$. Define $s_0:\rz_+\to\rz_+$ by
  \eqref{vollquad}, set $\xi:=\max\{X(t)/t^3|t\in\rz_+\}/(4\pi)$, and
  write $-e^\mathrm{TF}$ for the Thomas-Fermi energy of
  hydrogen. Then, for all $\rho\in P$ with
  $\int_{\rz^3}\rd x \rho(x)=N$,
  \begin{equation}
    \label{schrank}
    \TFW(\rho)\geq -{4s_0(\kappa)^5\over5\tf(s_0(\kappa))}e^\mathrm{TF}Z^\frac73 -\xi cN.
  \end{equation}
  Moreover, $s_0$ is strictly monotone increasing with $s_0(0)=0$ and
  $s_0(\kappa)\to\infty$ as $\kappa\to\infty$.
\end{satz}

\section{Application on the Excess Charge Problem\label{Ueberschuss}}
In this section we will show that the bound \eqref{resultat} allows
for an adaptation of the ideas of \cite{Franketal2018,Franketal2018T}
and show a bound on the excess charge of the relativistic
Thomas-Fermi-Weizs\"acker-Dirac atom which complements the bound
obtained in \cite{ChenSiedentop2020} in the absence of the exchange
term.

We define a monotone increasing function $\alpha:\rz\to[0,\pi/2]$ by
\begin{equation}
  \label{alpha}
  \alpha(s):=
  \begin{cases}
    0 & s\leq0\\
    \frac\pi2s &s\in(0,1)\\
    \frac\pi2& s\geq1
  \end{cases}.
\end{equation}
We introduce two localization functions
\begin{equation}
  \label{L}
  R:=    \sin\circ\alpha\ \mathrm{and}\ L:=\cos\circ\alpha,
\end{equation}
and corresponding localization functions $U$ and $O$ on $\rz^3$
defined by
\begin{equation}
  \label{UO}
  U(x):= L\left({\omega\cdot x -l\over s}\right),\ O(x):= R\left({\omega\cdot x -l\over s}\right)
\end{equation}
with the parameters $\omega\in\bS$, $l\in\rz_+$, and
$s\in(0,\infty)$. For later use, we write $A:=\supp(\nabla U)$ for the
support of the gradient of $U$ and $O$.

Assume $\rho_N$, with associated $p_N:=(3\pi^2\rho_N)^\frac13$, is a
minimizer of $\TFW$ under the constraint
\begin{equation}
  \label{neben=}
  \int_{\rz^3}\rho(x)\rd x =N.
\end{equation}
In abuse of notation, we sometimes write the occurring energy
functionals instead of depending on $\rho$ as depending on $p$, i.e.,
$p$ instead of $p^3/(3\pi^2)$.

Our starting point is the binding condition following directly from
the variational principle by pushing the $O$-part away from the
$U$-part
\begin{equation}
  \label{Bindung}
  \TFW(U p_N)+\cE^\mathrm{TFWD}_0(O p_N)-\TFW(p_N)\geq 0
\end{equation}
which is true, since
\begin{equation}
  \begin{split}
  &{1\over3\pi^2}\int_{\rz^3}(U(x)^3+O(x)^3)p_N(x)^3\rd x \leq {1\over3\pi^2}\int_{\rz^3}(U(x)^2+O(x)^2)p_N(x)^3\rd x\\
  = &\int_{\rz^3}\rho_N(x)\rd x =N
  \end{split}
\end{equation}
and the infima under the constraint \eqref{neben=} and the constraint
\begin{equation}
  \label{neben/=}
  3\pi^2\int_{\rz^3}p(x)^3\rd x =\int_{\rz^3}\rho(x)\rd x \leq N
\end{equation}
agree by \cite[Section 3.5]{Chen2019}. The corresponding argument,
namely pushing the charge difference between $N$ and the charge of the
minimizer to infinity, is a standard argument and works also when the
Dirac term is included.

We also have by the product rule
\begin{equation}
  \label{25}
  \begin{split}
    &\int_{\rz^3}\rd x[ |\nabla(Up)(x)|^2f(Up(x)/c)^2+|\nabla(Op)(x)|^2f(Op(x)/c)^2 ]\\
    \leq  &\int_{\rz^3}\rd x[ |\nabla(Up)(x)|^2+|\nabla(Op)(x)|^2]f(p(x)/c)^2 \\
    = &\int_{\rz^3}\rd x |\nabla p(x)|^2f(p(x)/c)^2
    +\int_{\rz^3}\rd x p(x)^2[|\nabla U(x)|^2+|\nabla O(x)|^2]f(p(x)/c)^2\\
    =&\int_{\rz^3}\rd x |\nabla p(x)|^2f(p(x)/c)^2+s^{-2}\int_{\rz^3}\rd
    x \alpha'((\omega\cdot x-l)/s)^2p(x)^2f(p(x)/c)^2,
  \end{split}
\end{equation}
since $f$ is monotone increasing.

An elementary calculation shows that there is a constant $\mu$ such
for all $t\in\rz_+$
\begin{equation}
  \label{mu}
  f(t)^2\leq \mu t.
\end{equation}
The optimal constant, namely $\max\{f(t)^2/t|t>0\}$, is
$\mu\approx1.66$ achieved at $t\approx1.45$.

For $\alpha\in\rz_+$ and $x\in\rz^3$ we claim
\begin{equation}
  \label{17}
    \int_\bS{\rd\omega\over4\pi}(\omega\cdot
    x-\alpha)_+=\frac{|x|}4[(1-{\alpha\over|x|})_+]^2
\end{equation}
(see \cite{Franketal2018} for a related formula). Since the left side
is independent of the direction of $x$ and equals
$|x| \int_\bS \rd \omega (4\pi)^{-1}(\omega\cdot x/|x|-\alpha/|x|)$,
it suffices to show \eqref{17} for $x=\ge_3$:
\begin{equation*}
  \begin{split}
    &\int_\bS{\rd\omega\over4\pi}(\omega\cdot\ge_3-\alpha)_+
    = \tfrac12\int_0^\pi\rd\vartheta\sin\vartheta(cos\vartheta-\alpha)_+
    = \tfrac12\int_{\min\{1,\alpha\}}^1\rd u (u-\alpha)\\
    =&\tfrac14[(1-\alpha)_+]^2.
  \end{split}
\end{equation*}

We estimate the various parts of \eqref{Bindung} separately. We begin with
the Weizs\"acker terms and get using \eqref{25} and \eqref{mu}
\begin{equation}
  \label{BindungW}
  \begin{split}
    &\W(Up_N)+\W(Op_N)-\W(p_N)\\
    \leq&{3\lambda\over8\pi^2s^2} \int_{0<\omega\cdot x - l< s}\rd x\, \alpha'((\omega\cdot x-l)/s)^2p_N(x)^2cf(p_N(x)/c)^2\\
    =&{3\lambda\over32s^2} \int_{0<\omega\cdot x - l< s}\rd x\,
    p_N(x)^2cf(p_N(x)/c)^2\leq {3\lambda\mu\over32s^2}
    \int_{0<\omega\cdot x - l< s}\rd x\, p_N(x)^3\\
  = &{9\pi^2\lambda\mu\over32s^2}
  \int_{0<\omega\cdot x - l< s}\rd x\, \rho_N(x).
\end{split}
\end{equation}
Integration over $l\in\rz_+$ and $\omega\in\bS$ and using \eqref{17}
yields
\begin{equation}
  \label{Wa}
  \begin{split}
    &\int_\bS{\rd \omega\over4\pi} \int_{\rz_+}\rd l \left(\W(Up_N)+\W(Op_N)-\W(p_N)\right)\\
    \leq& {9\pi^2\lambda\mu\over128s^2} \int_\bS{\rd\omega\over\pi}
    \int_{\rz_+}\rd x\, (\omega\cdot x - (\omega\cdot x -s)_+ )_+
    \rho_N(x)\\
    =& {9\pi^2\lambda\mu\over128s^2}
    \int_\bS{\rd\omega\over\pi}\int_{\rz_+}\rd x [(\omega\cdot x)_+ -
    (\omega\cdot x -s)_+ ]
    \rho_N(x)\\
    = &{9\pi^2\lambda\mu\over128s^2}\int_{\rz_+}\rd
    x\,|x|[1-((1-{s\over|x|})_+)^2]\rho_N(x)\\
    = &{9\pi^2\lambda\mu\over128s^2} \left(\int_{s<|x|}\rd x \left(
        2s-{s^2\over|x|}\right)\rho_N(x) +\int_{s>|x|}\rd x
      |x|\rho_N(x)\right) \leq {3^2\pi^2\lambda\mu\over2^6}\frac Ns.
  \end{split}
\end{equation}

Next we estimate the combined Thomas-Fermi-Exchange part of
\eqref{Bindung}. To this end we introduce the functions $a$ and $b$ on
$\rz_+$ defined by
\begin{align}
  \label{a}
  a(t):= & {c^5\over 8\pi^2}\tf(t) + {c^4\over8\pi^3}3[t(t^2+1)^\frac12-\arsinh(t)]^2.\\
  b(t):= & {c^4\over8\pi^3}2t^4.        
\end{align}
Pick now $f_1,...,f_n\in\rz_+$ with
$f_1^2+...+f_n^2=1$. Since $a,...,a^{(iv)}$ are all positive, we have
\begin{equation}
  \label{3}
  a'''(f_it)\leq a'''(t)
\end{equation}
because $f_i\leq1$. Since also $a(0)=a'(0)=a''(0)=a'''(0)$,
integration of \eqref{3} yields successively
$a''(f_\nu t)\leq f_\nu a''(t)$, $a'(f_\nu t)\leq f_\nu^2a'(t)$,
and $a(f_\nu t)\leq f_\nu^3a(t)$. (See \cite[Formula 3.135]{Chen2019}
for a similar argument for $\tf(t)$.)

Thus, we get
\begin{equation}
  \label{BindungTF}
  \begin{split}
    &\TF(Up_N)-\X(Up_N)+\TF(Op_N)-\X(Op_N)-(\TF(p_N)-\X(p_N))\\
    =&\int_{\rz^3}\rd x[ a(U(x)\tfrac{p_N(x)}c)+a(O(x)\tfrac{p_N(x)}c)-a(\tfrac{p_N(x)}c)\\
    & +b(\tfrac{p_N(x)}c)-b(U(x)\tfrac{p_N(x)}c)-b(O(x)\tfrac{p_N(x)}c)]\\
    \leq& \int_{\rz^3}\rd x[(U(x)^3+O(x)^3-1)a(\tfrac{p_N(x)}c) -(U(x)^4+O(x)^4-1)
    b(\tfrac{p_N(x)}c)]\\
    \leq &{1\over4}\underbrace{\max\left\{{(1-\cos(t)^4-\sin(t)^4)^2\over 1-\cos(t)^3-\sin(t)^3}\Big| t\in[0,\pi/2]\right\}}_{=(2+\sqrt2)/4} \int_A\rd x {b(\tfrac{p_N(x)}c)^2\over a(\tfrac{p_N(x)}c)}\\
    &\leq {2+\sqrt2\over2^5} {c^3\over \pi^4} \int_A\rd x { (\tfrac{p_N(x)}c)^8\over \tf(\tfrac{p_N(x)}c)}.
  \end{split}
\end{equation}
Using \eqref{tfung1} and \eqref{tfung2} we get for any $S\in\rz_+$
\begin{equation}
  \label{BindungTFa}
  \begin{split}
    &\TF(Up_N)-\X(Up_N)+\TF(Op_N)-\X(Op_N)\\
    &-(\TF(p_N)-\X(p_N))\\
    \leq&{2+\sqrt2\over2^5\pi^4}\int_{A, p_N(x)/c<S}\rd x{S^5\over
      \tf(S)}p_N(x)^3\\
    &+ {2+\sqrt2\over2^5\pi^4}\int_{A, p_N(x)/c\geq S}\rd
    x{S^4\over c\tf(S)}p_N(x)^4\\
    \leq &{(2+\sqrt2)3\over2^5\pi^2} {S^5\over \tf(S)}N +{(2+\sqrt2)3^\frac43\over2^5\pi^\frac43c} {S^4\over \tf(S)}\int_{\rz^3}\rd x \rho_N(x)^\frac43.
  \end{split}
\end{equation}

The external potential part yields
\begin{equation}
  \label{ae}
  -Z \int_{\rz^3}\rd x {(U(x)^3-1)\rho_N(x)\over|x|}
  \leq Z\int_{\omega\cdot x- l>0}\rd x\, {\rho_N(x)\over|x|}.
\end{equation}

Integration over $l$ and averaging over the sphere yields
\begin{equation}
  \label{aea}
  \begin{split}
    &-  Z \int_{\bS}{\rd\omega\over4\pi}\int_{\rz_+}\rd l\int_{\rz^3}\rd x
    {(U(x)^3-1)\rho_N(x)\over|x|}\\
    \leq &Z  \int_{\bS}{\rd\omega\over4\pi}\int_{\rz_+}\rd l\int_{\omega\cdot x- l>0}\rd x\, {\rho_N(x)\over|x|}\\
    =&Z  \int_{\bS}{\rd\omega\over4\pi}\int_{\rz^3}\rd x\, {(\omega\cdot
      x)_+ \rho_N(x)\over|x|} =\frac Z4  \int_{\rz^3}\rd x\, \rho_N(x)=\frac Z4 N.
  \end{split}
\end{equation}

Finally, we address the electron-electron repulsion in
\eqref{Bindung}. We have
\begin{equation}
%  \label{D}
  \begin{split}
    W(l,\omega):=&D(U^3\rho_N,U^3\rho_N)+D(O^3\rho_N,O^3\rho_N)\\
    &-D((U^2+O^2)\rho_N,(U^2+O^2)\rho_N)\\
    \leq &-2D(U^2\rho_N,O^2\rho_N)
    \leq -\int_{\omega\cdot x -l<0}\rd x \int_{\omega\cdot y-l>s}{\rho_N(x)\rho_N(y)\over|x-y|}.
  \end{split}
\end{equation}
Integration in $l$ and $\omega$ and using \eqref{17} yields
\begin{equation}
  \label{Da}
  \begin{split}
    &\int_{\bS}{\rd\omega\over4\pi} \int_{\rz_+}\rd l W(l,\omega)
    \leq -\int_{\bS}{\rd\omega\over4\pi} \int_{\rz_+}\rd l\int_{\omega\cdot x -l<0}\rd x \int_{\omega\cdot y-l>s}\rd y{\rho_N(x)\rho_N(y)\over|x-y|}\\
    =&-\int_{\bS}{\rd\omega\over8\pi} \int_{\rz_+}\rd l
    \int_{\rz^3}\rd x \int_{\rz^3}\rd
    y\\
    &[\theta(l-\omega\cdot x)\theta(\omega\cdot y -s-l)+
    \theta(l-(-\omega\cdot y))\theta(-\omega\cdot x-s-l)]
    {\rho_N(x)\rho_N(y)\over|x-y|}\\
    =& -\int_{\bS}{\rd\omega\over8\pi} \int_{\rz^3}\rd x \int_{\rz^3}\rd
    y 
    {(\omega\cdot(y-x) - s)_+ \rho_N(x)\rho_N(y)\over|x-y|}.
  \end{split}
\end{equation}
Thus, with \eqref{17},
\begin{equation}
  \label{Db}
  \begin{split}
    &\int_{\bS}{\rd\omega\over4\pi} \int_{\rz_+}\rd l\, W(l,\omega)
     \leq -\frac18\int_{\rz^3}\rd x \int_{\rz^3}\rd
    y 
    \rho_N(x)\rho_N(y)\left[\left(1-{s\over|x-y|}\right)_+\right]^2\\
    =&-{N^2\over8} + \frac18\int_{\rz^3}\rd x \int_{\rz^3}\rd
    y 
    \rho_N(x)\rho_N(y)\left\{1-\left[\left(1-{s\over|x-y|}\right)_+\right]^2\right\}\\
     =&-{N^2\over8} + \frac18\int_{\rz^3}\rd x \int_{\rz^3}\rd
    y 
    \rho_N(x)\rho_N(y)
    \times
    \begin{cases} 1 &s\geq |x-y|\\
      {2s\over|x-y|}-\left({s\over|x-y|}\right)^2&s<|x-y|
    \end{cases}\\
    &\leq -{N^2\over8} +\frac s2\D[\rho_N].
  \end{split}
\end{equation}

Inserting these estimates in \eqref{Bindung} gives
\begin{equation}
  \label{add0}
  {3^2\pi^2\lambda\mu\over2^6}\frac Ns +{ZN\over4} -{N^2\over8} + c_1(S)N +{c_2(S)\over c} \int_{\rz^3}\rd x \rho_N(x)^\frac43 +\frac s2\D[\rho_N] \geq0
\end{equation}
or
\begin{equation}
  \label{add}
  \begin{split}
    N\leq 2Z +
    {3^2\pi^2\lambda\mu\over 2^3s}+2^2s{\D[\rho_N]\over N} +8c_1(S) +{8c_2(S)\over cN} \int_{\rz^3}\rd x \rho_N(x)^\frac43
  \end{split}
\end{equation}
and after optimization in $s$
\begin{equation}
  \label{ergeb}
  N\leq 2Z +3\pi\sqrt2\sqrt{{\lambda\mu \D[\rho_N]\over N}} +8c_1(S) +{8c_2(S)\over cN} \int_{\rz^3}\rd x \rho_N(x)^\frac43.
\end{equation}

Now, we assume $\kappa=Z/(c\sqrt\lambda)$ fixed and apply Theorem
\ref{satz1} with a factor 2 in front of the exchange term and $Z$
replaced by $2Z$.  This gives \eqref{schrank} but with the
corresponding replacements, namely $Z$ by $2Z$ and a factor 2 in
front of $\xi cN$. Therefore we get
\begin{equation}
  \begin{split}
  0\geq& \TFW(\rho_N) 
  = \frac12\W(\rho_N) +\frac12 \TF(\rho_n) +\frac12\D[\rho_N]\\
  &+  \frac12\left(\W(\rho_N) +\TF(\rho_N) +\D[\rho_N]
    -\int_{\rz^3}\rd x {Z\rho_N(x)\over2|x|}-2\X(\rho_N)\right)\\
    \geq &\frac12\W(\rho_N) +\frac12 \TF(\rho_N) +\frac12\D[\rho_N]
    - C_\kappa Z^{7/3}-\xi cN.
    \end{split}
\end{equation}
Thus, all three terms, $\W(\rho_N)$, $\TF(\rho_N)$, and $\D(\rho_N)$
are bounded by a constant times $Z^{7/3}+cN$.

Now, $\tf(t)\geq 2t^4-(8/3)t^3$. Thus
\begin{equation}
  \begin{split}
  &\int_{\rz^3}\rho_N(x)^\frac43=
  {c^4\over(2\pi^2)^\frac43}\int_{\rz^3}\rd x\left(p_N(x)\over c\right)^4\\
  \leq & \left(\tfrac2\pi\right)^\frac23\left(c^{-1}\TF(\rho_N)+ c N\right)
  \leq D_\kappa(Z^\frac43\lambda^\frac12+N+cN)
  \end{split}
\end{equation}
with a $\kappa$-dependent constant $D_\kappa$.  Thus, \eqref{ergeb}
yields the following bound on the excess charge.
\begin{satz}
  \label{satz2}
  Assume that $\rho\in P$ with $\TFW(\rho)=\inf\TFW(P)$, set
  $N:=\int_{\rz^3}\rho(x) \rd x$, and assume $\kappa$ and $\lambda$
  positive and fixed. Then, for large $Z$,
\begin{equation}
  \label{ergebnise}
  N\leq 2Z+ O(Z^\frac23).
\end{equation}
\end{satz}
This should be compared to the bound $ N\leq 2.56 Z$ of \cite[Formula
(18)]{ChenSiedentop2020} for the relativistic
Thomas-Fermi-Weizs\"acker functional without exchange, i.e., even with
exchange term included we are lead to an improved leading order. Note,
however, it comes at a price, namely the ratio $Z/c$ and $\lambda$ is
now fixed.

\section*{Acknowledgments}
Partial support by the U.S. National Science Foundation through grants
DMS-1363432 and DMS-1954995 (R.L.F.) and by the Deutsche
Forschungsgemeinschaft (DFG, German Research Foundation) through
Germany's Excellence Strategy  EXC - 2111 - 390814868 (R.L.F., H.S.)
is gratefully acknowledged.
H.C. and H.S. thank the Institute of Mathematical Sciences at the
National University of Singapore for support through the program
\textit{Density Functionals for Many-Particle Systems: Mathematical
  Theory and Physical Applications of Effective Equations} which
inspired this work.

%\bibliographystyle{plain}
%\bibliography{coulomb}

\begin{thebibliography}{10}

\bibitem{Chen2019}
Hongshuo Chen.
\newblock {\em On the Excess Charge Problem in Relativistic Quantum Mechanics}.
\newblock PhD thesis, Ludwig-Maximilians-Universit{\"a}t M{\"u}nchen, July
  2019.

\bibitem{ChenSiedentop2020}
Hongshuo Chen and Heinz Siedentop.
\newblock On the excess charge of a relativistic statistical model of molecules
  with an inhomogeneity correction.
\newblock {\em Journal of Physics A: Mathematical and Theoretical},
  53(39):395201, September 2020.

\bibitem{EngelDreizler1987}
E.~Engel and R.~M. Dreizler.
\newblock Field-theoretical approach to a relativistic
  {T}homas-{F}ermi-{D}irac-{W}eizs\"acker model.
\newblock {\em Phys. Rev. A}, 35:3607--3618, May 1987.

\bibitem{Franketal2018}
Rupert~L Frank, Phan~Th{\`a}nh Nam, and Hanne Van Den~Bosch.
\newblock The ionization conjecture in {T}homas-{F}ermi-{D}irac-von
  {W}eizs{\"a}cker theory.
\newblock {\em Communications on Pure and Applied Mathematics}, 71(3):577--614,
  2018.

\bibitem{Franketal2018T}
Rupert~L. Frank, Phan~Th\`anh Nam, and Hanne Van Den~Bosch.
\newblock The maximal excess charge in {M}\"{u}ller density-matrix-functional
  theory.
\newblock {\em Ann. Henri Poincar\'{e}}, 19(9):2839--2867, 2018.

\bibitem{Gombas1949}
P.~Gomb{\'a}s.
\newblock {\em Die statistische {T}heorie des {A}toms und ihre {A}nwendungen}.
\newblock Springer-Verlag, Wien, 1 edition, 1949.

\bibitem{Gombas1956}
P.\ Gomb\'as.
\newblock {S}tatistische {B}ehandlung des {A}toms.
\newblock In S.\ {Fl\"ugge}, editor, {\em {H}andbuch der {P}ysik. Atome II},
  volume~36, pages 109--231. Springer-Verlag, Berlin, 1956.

\bibitem{Lieb1982A}
Elliott~H. Lieb.
\newblock Analysis of the {T}homas-{F}ermi-von {W}eizs\"{a}cker equation for an
  infinite atom without electron repulsion.
\newblock {\em Comm. Math. Phys.}, 85(1):15--25, 1982.

\bibitem{LiebLiberman1982}
Elliott~H. Lieb and David~A. Liberman.
\newblock Numerical calculation of the {T}homas-{F}ermi-von {W}eizs\"acker
  function for an infinite atom without electron repulsion.
\newblock Technical Report LA-9186-MS, Los Alamos National Laboratory, Los
  Alamos, New Mexico, April 1982.

\bibitem{Weizsacker1935}
C.~F. v.~Weizs{\"a}cker.
\newblock Zur {T}heorie der {K}ernmassen.
\newblock {\em Z.\ Phys.}, 96:431--458, 1935.

\bibitem{YoneiTomishima1965}
Katsumi Yonei and Yasuo Tomishima.
\newblock On the {W}eizs{\"a}cker correction to the {T}homas-{F}ermi theory of
  the atom.
\newblock {\em Journal of the Physical Society of Japan}, 20(6):1051--1057,
  1965.

\end{thebibliography}
\def\cprime{$'$}

\end{document}